\documentclass{article}    
\usepackage{amsmath,amssymb}

\newcommand\tit[1]{``#1,''}
\newcommand\pp[2]{#1\,--\,#2}
\newcommand{\R}{\mathbb{R}}
\newcommand{\C}{\mathbb{C}}

\newcommand{\Z}{\mathbb{Z}}
\newcommand{\Mat}[2]{\left(\begin{array}{#1}#2\end{array}\right)}
\newcommand{\mx}[1]{\left(\begin{smallmatrix}#1\end{smallmatrix}\right)}
\newcommand\Cl{\mathfrak{Cl}}
\newcommand{\e}{\mathbf e}
\newcommand{\f}{\mathbf f}
\newcommand{\Hil}{{\mathcal H}}
\newcommand{\ket}[1]{| #1 \rangle}
\newcommand{\bra}[1]{\langle #1 |}
\newcommand{\ktbr}[2]{\ket{#1}\bra{#2}}

\newcommand\bsig{\boldsymbol{\sigma}}
\newcommand{\Eq}[1]{{\rm Eq.~(\ref{#1})}}
\newcommand{\Sec}[1]{{\rm Sec.~\ref{Sec:#1}}}
\newcommand{\op}[1]{\boldsymbol{#1}}
\newcommand\btau{\boldsymbol{\tau}}
\newcommand{\Rf}{\varXi}

\title{From Fermat's Last Theorem\\
To the Quantum Computer}
\author{Alexander Yu.\ Vlasov\\
\small Federal Radiological Center, IRH\\
\small 197101, Mira Street 8, St.--Petersburg, Russia}
\date{3 July 2003}
\begin{document}
\sloppy
\maketitle                 
\begin{abstract}
Despite of an active work of many researchers in the theory of quantum
computations, this area still saves some mysterious charm. It is
already an almost common idea, that maybe many fashionable current
projects will fade in future, but some absolutely unpredictable
applications appear instead. Why such optimistic predictions are
legal here, despite of an extreme difficulty to suggest each one new
promising quantum algorithm or realistic ``industrial'' application?
One reason --- is very deep contents of this area. It maybe only an
extremely unlucky occasion, if such a fundamental thing won't supply us
with some bright insights and serious new applications. A sign of such
nontrivial contents of a theory --- are unexpected links between different
branches of our knowledge. In the present paper is mentioned one such link
--- between application of Weyl quantization in the theory of quantum
computations and abstract mathematical constructions born in mid of XIX
century due to unsuccessful tries to prove Fermat's last theorem.

\noindent\copyright~{\em A. Yu. Vlasov, 2003}

\noindent\copyright~{\em Nova Science Publishers, Inc., 2003}
\end{abstract}

\section{Introduction}
\label{Sec:intro}

The title of the paper is partially inspired by an old lecture presented
at 1929 in Liverpool by John E. Littlewood and reprinted in \cite{Little}.
Let us briefly recollect ideas of this work: unsuccessful
tries to prove Fermat's last theorem encourage introduction of new ideas
of {\em special abstract classes} of numbers: {\em ``ideals''} and
generalization of such ideas to other areas of mathematics.

 From the point of view of Littlewood some constructions developed during such
tries have encouraged to introduce a mathematical notion of a function as
some class. He criticized an ``old'' idea to consider a function $y = y(x)$
as some {\em method of calculation} of a value $y$ of the function from an
argument $x$ using {\em series of operations} instead of the ``new'' idea to
define the function simply as {\em a class of ordered pairs $(x,y)$ where all
$x$ are different}.

One question discussed in the mentioned lecture was the specific determinism
related with such definition of a function: if an evolution of any system
instead of $S = S(t)$ is treated as a class of pairs $(S,t)$, then such a
model already does not look like an ``evolution'' and rather resembles
a ``historical record'' or adopted to description to some mechanical
deterministic evolution without any options.

Of course the definition of a function as a class of pair is a standard thing
then and now, but the ``old'' definition should not be rejected, because in
works of Turing, Post and Church written just few years after this
Littlewood's lecture the ``old'' idea of a function and ``series of
operations'' was used as a basis of {\em the theory of recursion} and later
it was developed to the modern computer science. Let us denote such
definition {\em an operational}.

The operational approach also related with the description of some
``stand-alone'' function (c.f. {\em algorithm}) {\em vs} a definition using
the set of arguments and the values.  It also raises some new questions hardly
expressed in another approach.  One such question is {\em universality}.

If we are trying to express functions as a set of operations, it is reasonable
to consider a question, if the operations are powerful enough. For the definition
with a set of pairs, such a question is some kind of tautology, because it is
possible to choose any class of pairs we want, and it could be considered
as yet another demonstration of an elegance of the ``new'' definition, but the problem
is that such a definition often could hardly be applied to the real world. It is
especially clear for functions with an infinite (or very big finite) domain,
then instead of a short string like $y(x) = x^2$ it is suggested to consider
a table like
$\{(0,0),(1,1),(2,4),(3,9),(4,16), \ldots \}.$

 From such a constructive point of view the operational definition of a function
is preferable and so the question about an accessible set of basic operations is
reasonable. Here the set of basic operations is considered as an universal, if it is
possible to represent any function. For the infinite set of arguments it is
more difficult to explain such idea, but for the purposes of given paper
it is enough to consider finite sets.

It is also useful to consider a physical analogue of such a question:
is it possible to suggest some set of an elementary universal operations
for modeling of the arbitrary physical process? Such an idea was discussed
for example by Richard Feynman at 1981 in his lecture at PhysComp'81
conference in MIT \cite{FeySim}.

The quantum mechanics is a fundamental theory about our world, so it was
reasonable to give an answer about the universal set of operations using some
simple quantum mechanical models \cite{FeySim,FeyComp,DeuTur,DeuNet}.

It is interesting, that some constructions used in the theory of
quantum computation
also have a close relation with the algebraic ideas developed due to
the tries to prove Fermat's last theorem at mid XIX century and
discussed in Littelwood's work. These ideas are briefly recollected
in \Sec{alg} together with related algebraic constructions like the
group algebras, Clifford algebras, etc. In \Sec{univ} it is discussed
a relation between Lie algebras and universality in quantum computation.
It is used for a special construction of the universal set of gates using
Lie and Clifford algebras discussed in \Sec{qubit}. Such universal
elements may be applied to an array of two-dimensional quantum systems,
{\em the qubits}. It has an analogue with the binary logic and arithmetic.
An application of the similar ideas of universality to the higher-dimensional
quantum system based on ideas of Weyl quantization is represented
in \Sec{Weyl}. Constructions used in this section are close related
with algebraic ideas discussed in \Sec{alg}. In \Sec{QC} are discussed
some other areas of quantum computation linked with such an algebraic
approach to the discrete mathematics.

\section{Complex, algebraic, Clifford numbers and all that}
\label{Sec:alg}

Fermat's last theorem declares impossibility to resolve an equation
\mbox{$x^l+y^l=z^l$} for natural numbers $X,Y,Z$ and $l > 2$. For $l=2$ there
are infinite amount of natural solutions of the equation. For example
it is possible to choose $X=a^2-b^2$, $Y=2ab$, $Z=a^2+b^2$ for an arbitrary
natural $a>b$. It may be checked directly, but more useful to derive those
expressions with application of {\em complex numbers}.

Let $z=a+b\,i$ is a complex number and $\|z\|=a^2+b^2$ is the square of the
norm. Because $\|z^2\| = \|z\|^2$ and $z^2 = (a^2-b^2)+2ab\,i$, it is possible
to rewrite it as $(a^2+b^2)^2 = (a^2-b^2)^2 + (2ab)^2$ and it corresponds
to the definitions of $X,Y$ and $Z$ used above.

Let us not discuss in details neither the history of proof of Fermat's
last theorem \cite{FermPop}, nor the theory of algebraic numbers
\cite{FermPost} and simply describe some useful constructions introduced
due to these tries.

It is possible to write the norm of a complex number as $a^2+b^2=(a+b\,i)(a-b\,i)$,
but for an arbitrary sum of two odd powers ($l=2k+1$) it is
possible to write (again using complex numbers)
\begin{equation}
a^l + b^l = (a+b)(a+\zeta\,b)(a+\zeta^2 b)\cdots(a+\zeta^{l-1}b),
\quad \zeta = \sqrt[l]{-1} = e^{2 \pi i/l},
\label{LameProd}
\end{equation}
it follows from a more general solution for {\em arbitrary} $l$
\begin{equation}
 a^l - b^l = (a-b)(a-\zeta\,b)(a-\zeta^2 b)\cdots(a-\zeta^{l-1}b),
\quad \zeta = \sqrt[l]{-1} = e^{2 \pi i/l}.
\end{equation}
It may be rewritten as
\begin{equation}
 a^l +(-1)^{l-1} b^l = (a+b)(a+\zeta\,b)(a+\zeta^2 b)\cdots(a+\zeta^{l-1}b),
\end{equation}
or
\begin{equation}
 a^l + b^l = (a-\nu b)(a-\nu^3\,b)(a-\nu^5 b)\cdots(a-\nu^{2l-1}b),
\quad \nu = \sqrt{\zeta} = e^{\pi i/l},
\end{equation}
using the substitutions $b \to -b$ or $b \to \nu b$ respectively.

It was already discussed above, how ``complex-integer'' numbers like
$a+b\,i$, $a,b \in \Z$ may help with solution of a quadratic equation
with integer coefficients, but it is useful also to introduce more general
{\em algebraic integer} numbers like $\sum a_k \zeta^k$ \cite{FermPost}.

At 1843 E. Kummer and 1847 G. Lam\'e suggested to use such numbers for a proof
of Fermat's last theorem for any prime power $l$. It was generalizations of
Euler's proof for $l=3$ and, roughly, an idea was related with \Eq{LameProd}
describing two different decompositions of the same number (as the power of
$z$ and as the product \Eq{LameProd}), but such a thing is impossible for
{\em usual natural numbers} there each number may be expressed as an unique
product of the prime numbers. The problem with such a proof was found soon by
P. Dirichlet, E. Kummer, J. Liouville and related with non-uniqueness of
decomposition of the sums $n_k \zeta^k$ introduced by Kummer and Lam\'e
($\zeta^l=1$) for some $l$ \cite{FermPost}.

Really there are some subtleties non-relevant for present consideration,
for example for any $l$: $1+\zeta+\cdots+\zeta^{l-1}=0$ and so
some sum vanishes. For prime $l$ it is enough to exclude
only $\zeta^{l-1}$, but here the theory of algebraic numbers is considered
only as some intermediate step.

Really, let us consider $\zeta$ not as some complex number, but as an element
of some abstract cyclic group $\Z_l$ generated by the powers of $\zeta$ with
the property $\zeta^l=1$. For any $l$ it is possible to consider {\em the
group algebra} described by formal series $\sum a_k \zeta^k$ of such elements
with naturally defined laws of addition and multiplication \cite{SLang}.
Example of representation of such algebra is the algebra of {\em diagonal} $l
\times l$ matrixes generated by matrix $V$ defined below by \Eq{WeylPair}.

The group algebra described above is a commutative algebra. Let us return
again to the case $l=2$. Earlier it was described the representation of a sum of
squares as $a^2+b^2=(a+b\,i)(a-b\,i)$, but it is also possible to rewrite
that using a representation of the imaginary unit as the real $2 \times 2$
matrix $i \leftrightarrow \mx{0&-1\\1&0}$ (realification) and so
the decomposition of $a^2+b^2$ may be considered as some matrix equation.

It is also possible to write the quadratic form not as
a product of two different terms, but as the full square using other
$2 \times 2$ matrices $\e_1 = \mx{-1&0\\0&1}$, $\e_2 = \mx{0&1\\1&0}$;
$a^2+b^2=(a\,\e_1 +b\,\e_2)^2$.
Such representation is possible because $\e_1^2=\e_2^2=1$,
$\e_1\e_2 = -\e_2\e_1$ and so \mbox{$(a\,\e_1 +b\,\e_2)^2 = a^2 + ab\,\e_1\e_2 +
ba\,\e_2\e_1 + b^2 = a^2+b^2$}.
It is already the noncommutative algebra.

A generalization of such an equation for the arbitrary number of terms
\begin{equation}
 (a_1 \e_1 + a_2 \e_2 + \cdots + a_n \e_n)^2 =
  a_1^2 + a_2^2 + \cdots + a_n^2
\end{equation}
is equivalent with the definition of Clifford algebra $\Cl(n)$
\cite{PostLie,ClDir}
\begin{equation}
 \e_i \e_j + \e_j \e_i = 2 \sigma_{ij}.
\label{ClPl}
\end{equation}

Similar noncommutative algebraic version of condition for Fermat's
last theorem is the equation
\begin{equation}
 (a_1 \f_1 + a_2 \f_2)^l = a_1^l + a_2^l,
\end{equation}
where $\f$ are elements of some noncommutative algebra.

Let us show, that two elements of an algebra with the property
\begin{equation}
 \f_1^l = \f_2^l = 1, \quad \f_1 \f_2 = \zeta \f_2 \f_1
 \quad (\zeta = e^{2 \pi i/l}),
\end{equation}
satisfy to the necessary equation. Really
\begin{eqnarray*}
 &&(a_1 \f_1 + a_2 \f_2)^l =
 (a_1 + a_2 \f_2\f_1^{-1})\f_1 (a_1 + a_2 \f_2\f_1^{-1})\f_1 \cdots
 (a_1 + a_2 \f_2\f_1^{-1})\f_1 \\
 && \quad= (a_1 + a_2 \f_2\f_1^{-1}) (a_1 + \zeta a_2 \f_2\f_1^{-1}) \cdots
 (a_1 + \zeta^{l-1}a_2 \f_2\f_1^{-1})\,(\f_1)^l \\
 && \quad = a_1^l + (-1)^{l-1}a_2^l (\f_2\f_1^{-1})^l = a_1^l + a_2^l.
\end{eqnarray*}

It is also possible to satisfy the equation
\begin{equation}
 (a_1 \f_1 + a_2 \f_2 + \cdots + a_n \f_n)^l =
  a_1^l + a_2^l + \cdots + a_n^l
\end{equation}
using the noncommutative algebra with $n$ generators $\f_i$ and
the relations \cite{WeylClif}
\begin{equation}
 \f_i^l = 1, \quad \f_i \f_j = \zeta \f_j \f_i,
 \quad i < j.
\label{zetcom}
\end{equation}

\section{Universal quantum gates and Lie algebras}
\label{Sec:univ}

Let us return to the theory of computation and universal sets of operations.
It was already mentioned an idea to apply a similar theory to the physical
systems and find some set of universal operations \cite{FeySim,FeyComp,%
DeuTur,DeuNet,DeuUn}.

Here is discussed a simple model with the quantum world
described as {\em a finite-dimensional Hilbert space} (the complex
vector space $\C^n$ with Hermitian scalar product) and unitary operators
on this space ($n \times n$ complex matrices $U$ with the property $U U^*=1$).
In the theory of quantum computations the matrices are called {\em the quantum
gates}.
It is also used an abstract operation of the composition of such systems
described as the tensor product $\C^n \otimes \C^m \cong \C^{mn}$.

For such a model the question about universality may be reformulated as
a necessity to find some set of unitary matrices (the quantum gates)
$\{U_\mu\}$ with possibility to express
any unitary transformation $U$ as a product of the matrices (gates) from
the set $U = U_{\mu_1} U_{\mu_2} \cdots U_{\mu_k}$. For a finite set
$\{U_\mu\}$ the index $\mu$ is simply a natural number. In such a case
it is impossible to represent any matrix $U$ {\em precisely} using a finite
number of terms $U_{\mu_k}$, but it is enough to consider the possibility to
{\em approximate} any matrix with an arbitrary accuracy
$U \approx U_{\mu_1} U_{\mu_2} \cdots U_{\mu_k}$
\cite{DeuTur,DeuNet,DeuUn}. Sometime it is called {\em the universality
in approximate sense}.

The group of unitary matrices is Lie (``smooth'') group. It was found,
that the Lie algebra of the Lie group is a convenient tool for the theory of
universality \cite{DeuUn,DV95}. The idea uses correspondence between
the operations like {\em addition} $\mathbf{a+b}$ and {\em Lie bracket}
$\mathbf{[a,b]}$
for elements of Lie algebras with operations $AB$ and $ABA^{-1}B^{-1}$
for Lie group.

So instead of elements $\{U_\mu\}$ of Lie group SU$(n)$ it is possible
to consider elements $\{\op u_\mu\}$ of Lie algebra su$(n)$ and the notion
of universality should be adopted to the Lie algebra.

\noindent~\textbullet~
{\em If a set of elements $\{\op u_\mu \in {\rm su}(n)\}$  may generate
the full algebra using additions and commutators, then the set is called
universal}.

Using such a set of elements and the map $U_\mu = \exp(\tau \op u_\mu)$
with small $\tau$ it is possible to construct the universal set
of operators $\{U_\mu\in {\rm SU}(n)\}$ \cite{DeuUn,DV95}.

It is also possible to consider $\tau$ as a continuous parameter
and to use the family of gates $\{U_\mu(\tau)\}$
for construction of the {\em ``strictly''} universal set.
It should be mentioned, that an element of Lie algebra corresponds to
Hamiltonian used for construction of the gate $\op H_\mu = i\,\op u_\mu$
and in such a case the parameter $\tau$ corresponds to the time.
It follows directly from the solution of the Schr\"odinger equation with
time-independent Hamiltonian
$$
 \dot\psi = -i \op H \psi \quad \Longrightarrow
 \quad \psi(t) = \exp(-i \op H t)\,\psi.
$$

\section{Qubits}
\label{Sec:qubit}

Let us consider a simplest example with two-dimensional Hilbert spaces
$\Hil_2$. Two vectors of a basis are usually denoted as $\ket{0}$ and
$\ket{1}$, i.e.
$$
 \ket{0} = \Mat{c}{1\\0}, \quad \ket{1} = \Mat{c}{0\\1}, \quad
 \alpha\ket{0}+\beta\ket{1} = \Mat{c}{\alpha\\ \beta} \in \Hil_2.
$$
Such an abstract quantum system with two states are usually called {\em the
qubit}. A quantum gate, i.e. an unitary matrix $U \in \mathrm U(2)$ may be
expressed as
$$
 U = e^{i\varphi} (a_0 + a_1 i\bsig_1 + a_2 i\bsig_2 + a_3 i\bsig_3),
\quad a_k \in \R,~~a_0^2+a_1^2+a_2^2+a_3^2 = 1,
$$
where
\begin{equation}
 \bsig_1 = \Mat{rr}{0&1\\1&0},\quad
 \bsig_2 = \Mat{rr}{0&-i\\i&0},\quad
 \bsig_3 = \Mat{rr}{1&0\\0&-1}
\label{PauliMat}
\end{equation}
are three Pauli matrices and $e^{i\varphi}$ is an unessential phase multiplier.

It can be checked directly, $\bsig_1$ corresponds to the classical ${\sf NOT}$
gate: $\bsig_1\ket{0}=\ket{1}$, $\bsig_1\ket{1}=\ket{0}$ and it explains the
idea to consider unitary matrices as analogues of classical gates.

It is convenient to use Pauli matrices also as
elements of the Lie algebra and there is the simple expression
\begin{equation}
 \exp(i \bsig_k \tau) = \cos(\tau) + i\bsig_k \sin(\tau),
 \quad k = 1,2,3.
\end{equation}

A quantum state of $n$ such systems is described as the tensor product
\begin{equation}
 \Hil = \underbrace{\Hil_2\otimes\cdots\otimes\Hil_2}_n
\label{Hil2prod}
\end{equation}
and has the dimension $2^n$.  Basic vectors for such states may be
denoted as $\ket{0\ldots 00},\ket{0\ldots 01},\ldots,\ket{1\ldots 11}$.
A general element of such Hilbert space may be written as
\begin{equation}
 \ket{v} =
 v_{0\ldots 00}\ket{0\ldots 00}+v_{0\ldots 01}\ket{0\ldots 01}+
 \cdots+v_{1\ldots 11}\ket{1\ldots 11}.
\end{equation}
It is convenient sometime to consider it as {\em a binary decomposition} of
indexes of vectors $v$ and elements of basis.

Quantum gates are described by $2^n \times 2^n$ unitary matrices
SU$(2^n)$ and the action of such matrix $\ket{v'}=U\ket{v}$ may be
rewritten as
\begin{equation}
 v'_{i_1 i_2 \ldots i_n} = \sum_{j_1,j_2,\ldots,j_n = 0}^1%
 {U^{j_1 j_2 \ldots j_n}_{i_1 i_2 \ldots i_n} v_{j_1 j_2 \ldots j_n}}.
\end{equation}
Here set of indexes like $i_1 i_2 \ldots i_n$ is again simpler to compare
with binary decomposition of some number.

But it is possible also to consider an action of some ``one-gate''
$U \in \mathrm U(2)$  on a singular qubit with index $k$
\begin{equation}
 v'_{i_1\ldots i_k \ldots i_n} = \sum_{j_k = 0}^1%
 {U^{j_k}_{i_k} v_{i_1\ldots j_k \ldots i_n}},
\end{equation}
or an action of some ``two-gate''
$U \in \mathrm U(4)$  on two qubits with indexes $k,l$
\begin{equation}
 v'_{i_1 \ldots i_k \ldots i_l \ldots i_n} = \sum_{j_k,j_l = 0}^1%
 {U^{j_k j_l}_{i_k i_l} v_{i_1\ldots j_k \ldots j_l \ldots i_n}},
\end{equation}
and similarly with any $k$-gate, $k\le n$.

It is also convenient to use the basis of $2^n \times 2^n$ complex matrices
expressed via $4^n$ different tensor products of Pauli matrices together
with the unit matrix $\bsig_0 \equiv \mx{1&0\\0&1}$ \cite{BR94,SCH98}
\begin{equation}
 \bsig_{i_1} \otimes \bsig_{i_2} \otimes \cdots \otimes \bsig_{i_n},
\quad i_1,i_2,\ldots,i_n = 0,\ldots,3.
\end{equation}
In such a notation the $k$-gate is a sum of terms with non-unit elements
$\bsig_i$ ($i=1,2,3$) only in $k$ {\em given} positions.

\medskip

Already for such a simple case with the qubits systems it is clear, that a
complexity of the models is very high, for example for the composition of
$10$ qubits Hilbert space has dimension $2^{10}=1024$, for $20$ qubits
--- $2^{20}=1048576$, etc. Dimension of the space of unitary matrices is even
bigger: $4^n$, e.g. for $20$ qubits such a matrix contains
$1048576 \times 1048576 = 1099511627776$ complex numbers.

So question about the universal set of elements is actual here. Very important
results here are related with universality for sets of {\em two-gates}
\cite{DeuUn,DV95,Gates95}.

Together with {\em proofs of existence} for such sets of gates it is useful
to know some {\em constructive} algorithms and have possibility to decompose
or approximate some matrix or estimate of complexity of some class of
gates. From such a point of view the method of construction of an universal
set of gates based on mechanical testing of completeness of the commutator
algebra may be not very convenient. It is more useful, then the universal
set of gates has some clear algebraic structure.

In \cite{VlaUCl} was suggested to use Clifford algebras for construction
of the universal sets of quantum gates. It is especially convenient due to
interesting and useful relation between the structure of Clifford algebra
\cite{ClDir} and the product operator formalism \cite{BR94,SCH98}.
Really, generators of Clifford algebra for even dimension $\Cl(2n)$
satisfying \Eq{ClPl} may be expressed using Pauli matrices as \cite{ClDir}
\begin{eqnarray}
 \e_{2k} & = &
  {\underbrace{\bsig_0\otimes\cdots\otimes \bsig_0}_{n-k-1}\,}\otimes
 \bsig_1\otimes\underbrace{\bsig_3\otimes\cdots\otimes\bsig_3}_k \, ,
 \nonumber\\
 \e_{2k+1} & = &
 {\underbrace{\bsig_0\otimes\cdots\otimes \bsig_0}_{n-k-1}\,}\otimes
 \bsig_2\otimes\underbrace{\bsig_3\otimes\cdots\otimes\bsig_3}_k \, ,
 \label{defE}
\end{eqnarray}
where $k = 0,\ldots,n-1$ and $\bsig_0$ is $2\times 2$ unit matrix.
Clifford algebra $\Cl(2n)$ coincides with the algebra of $2^n \times 2^n$
complex matrices, the expressions for generators \Eq{defE} have clear product
structure and it is convenient for theory of quantum computations
(both usual and fermionic case) \cite{VlaCl}.

The elements \Eq{defE} do not correspond to two-gates, but it is possible
to consider elements $\e_{j,j+1} \equiv [\e_j,\e_{j+1}] = 2\e_j\e_{j+1}$.
Such elements of ``second order''
correspond to one-gates for $j=2k$ and two-gates for $j=2k+1$ and
generate Lie algebra so$(2n)$ represented by all possible
bi-products $\e_{jk} \equiv \e_j\e_k = [\e_j,\e_k]/2$. It {\em does not}
produce an universal set of gates, because dimension of such an algebra is
$(2n-1)n < 4^n-1 = \dim{\mathrm{SU}(2^n)}$. Lie group corresponding to
such elements is isomorphic with Spin$(2n)$.

It should be mentioned, that it is enough to add only two extra gates
to produce an universal set: it should be one initial element $\e_j$ and
an arbitrary product with three or four elements, for example it may be
two elements $\e_0$ and $\e_0\e_1\e_2$ \cite{VlaUCl}.
Second element has a third (or fourth) order and this property is important
for discussions on the fermionic quantum computation \cite{VlaCl,BK00}, but
in the product operator representation both elements may be chosen as
one-gates:
$$
\e_0 = 1\otimes \cdots \otimes 1 \otimes \bsig_1, \quad
\e_0\e_1\e_2 = 1\otimes \cdots \otimes \bsig_1 \otimes 1.
$$

\section{Univeraslity for $l \ge 2$ and Weyl quantization}
\label{Sec:Weyl}

The theory of quantum gates represented here may be quite general and
common, but it seems lack of some habitual attributes of the quantum mechanics.
Where is Heisenberg uncertainty relation, the coordinates and momenta,
the wave-particle duality and all that? Of course here are represented
the discrete models, but for significant amount of qubits dimension of Hilbert
space becomes very big and so there is some hope to consider an
analogue of a continuous limit.

A very convenient tool for such a problem is Weyl representation of
Heisenberg commutation relations and some other methods related with Weyl
quantization. It is discussed below.

Let us first instead of two-dimensional Hilbert space consider
finite-dimensional one with arbitrary dimension $l \ge 2$ and denote
it as $\Hil_l$. It is possible without a big problem to generalize
the most properties described above. The compound systems may be described
as the $l^n$-dimensional tensor product like with \Eq{Hil2prod}
\begin{equation}
 \Hil = \underbrace{\Hil_l\otimes\cdots\otimes\Hil_l}_n,
\label{Hillprod}
\end{equation}
an action of a matrix $U \in \mathrm{SU}(l^n)$ may be represented as
\begin{equation}
 v'_{i_1 i_2 \ldots i_n} = \sum_{j_1,j_2,\ldots,j_n = 0}^{l-1}%
 {U^{j_1 j_2 \ldots j_n}_{i_1 i_2 \ldots i_n} v_{j_1 j_2 \ldots j_n}},
\end{equation}
and actions of $k$-gates for $k<n$ also may be described using
same formulae, as for qubits, but with indexes range $(0,\ldots,l-1)$
instead of $(0,1)$.

\smallskip

The more nontrivial thing is to introduce an analogue of Pauli matrices,
but it also exists and in addition provides some passage to the continuous
limit mentioned above.

Let us consider usual Heisenberg commutation relation ($\hbar=1$)
\begin{equation}
 [\op p,\op q] = \op p\op q -\op q\op p = -i,
\end{equation}
where $\op p,\op q$ are operators of momentum and coordinate. Let us
consider two families of operators
\begin{equation}
 \op U^{\alpha} = \exp(i \alpha \op p), \quad
 \op V^{\beta} = \exp(i \beta \op q).
\label{opUVdef}
\end{equation}
Using Campbell-Hausdorff formula for the formal operator series (for operators
with zero third-order commutators like for $\op p,\op q$)
$$\exp(\op a + \op b) = \exp(\op a)\exp(\op b)\exp(-\frac12[\op a,\op b]),$$
it is possible to write
\begin{equation}
\op U^{\alpha}\op V^{\beta} = \exp(i\alpha\beta)\op V^{\beta}\op U^{\alpha}.
\label{Wcom}
\end{equation}
It is {\em Weyl system} \cite{AQFT} or {\em Weyl representation of Heisenberg
commutation relations}.

Weyl relation \Eq{Wcom} is even more general, than Heisenberg one
\cite{ReS}, e.g. it is applied to compact operators instead of $\op p,\op q$
and so widely used in many areas of the quantum theory. But in the present
paper these relations are used, because they work for {\em finite dimensional
Hilbert spaces} and this application was discussed already in initial Weyl
work at 1927 (first English translation at 1931 \cite{WeylGQM}).

Let us find two $l\times l$ unitary matrices with the property similar with
\Eq{Wcom}
\begin{equation}
\op U\op V = \zeta\op V\op U.
\label{UVVU}
\end{equation}
It can be shown that such matrices are really exist for $\zeta^l=1$
and may be written as \cite{WeylGQM}
\begin{equation}
 U = \Mat{ccccc}{0&1&0&\ldots&0\\0&0&1&\ldots&0\\
 \vdots&\vdots&\vdots&\ddots&\vdots\\0&0&0&\ldots&1\\1&0&0&\ldots&0}\!,
 \quad
 V = \Mat{ccccc}{1&0&0&\ldots&0\\0&\zeta&0&\ldots&0 \\
 0&0&\zeta^2&\ldots&0 \\ \vdots&\vdots&\vdots&\ddots&\vdots\\
 0&0&0&\ldots&\zeta^{l-1}}\!.
\label{WeylPair}
\end{equation}
So Weyl relations work also for the finite-dimensional case.
Really the ``shift'' and ``clock'' matrices \Eq{WeylPair} were introduced
even earlier, at 1882--84 in works of J. J. Sylvester \cite{Zachos}.

It should be mentioned, that due to \Eq{UVVU} the matrices are satisfying to
``operator Fermat's theorem'' discussed above, i.e.
\begin{equation}
 (a\op V + b\op U)^l = a^l + b^l.
\end{equation}

The matrices widely used in the theory of quantum computation after reintroducing
for the theory of quantum error correction \cite{KnillQEC,GottQEC}. The matrices
also may be very useful for the theory of universal quantum gates for the higher
dimensional quantum systems \cite{VlaUNt}, there it has analogue with
the application of Clifford algebras discussed above.
For $l=2$ the matrices coincide with $\bsig_1$ and $\bsig_3$.
In papers about quantum computer applications the Weyl pair $U,V$ is often
called {\em generalized Pauli matrices} with yet another notation $X,Z$.

Really such approach corresponds to some discrete analogue of Weyl
quantization. Let us discuss it in more details. In Weyl quantization
\cite{WeylGQM,Fadd} any function $f(p,q)$ with two real arguments $p,q$
and with Fourier image $\tilde f(\alpha,\beta)$ described by expression
\begin{equation}
 f(p,q) = \iint\limits_{-\infty}^{~+\infty}{\exp(i\alpha p + i\beta q)
 \tilde f(\alpha,\beta)\, d\alpha\, d\beta}
\end{equation}
is associated with the operator $\op f$ defined as
\begin{equation}
 \op f = \iint\limits_{-\infty}^{~+\infty}{\exp(i\alpha \op p + i\beta \op q)
 \tilde f(\alpha,\beta)\, d\alpha\, d\beta}.
\end{equation}
Using Hausdorff formula and the definition \Eq{opUVdef} it can be rewritten as
\begin{equation}
 \op f = \iint\limits_{-\infty}^{~+\infty}{\exp(-i\alpha\beta/2)
 \tilde f(\alpha,\beta) \op U^\alpha \op V^\beta\, d\alpha\, d\beta}.
\label{WUVdec}
\end{equation}

For finite case, there the integrals should be changed to sums, the
\Eq{WUVdec} would correspond to a decomposition of some matrices as a
sum with product of different integer powers of matrices $U,V$ \Eq{WeylPair},
like $\sum_{k,j} f_{kj}U^{k}V^{j}$ (up to nonsignificant complex multiplier
like $\zeta^{kj/2}$). Such decomposition is really always exist and
unique, because matrices $U^{k}V^{j}$ $(k,j = 0,\cdots,l-1)$ produce
basis in space of all $l \times l$ complex matrices. It is also
possible to use natural norm on space of matrices
$
 \|A,B\| = \mathrm{Tr}(A B^*)/l
$
to make the basis orthonormal.

It should be only mentioned, that matrices used in such decomposition are not
Hermitian. It is some difference with two-dimensional case and Pauli
matrices. It is possible to use following method to resolve such problem: if
there is some matrix A, then it is possible to consider two Hermitian
matrices $A+A^*$ and $i(A-A^*)$ instead of it. So instead of Lie algebra
su$(l)$ it is possible to consider Lie algebra sl$(l,\C)$ of all complex
matrices with trace zero.

\begin{quote}
{\em Note}:~
There is yet another way to represent Hermitian matrix using basis generated
by Weyl pair. Instead of $U^j V^k$ it is possible to use
{\em ``$90^\circ$ rotated''} matrices \cite{MPS,Paz}
\begin{equation}
 \zeta^{kj/2}U^j V^k \Rf,
\label{Ajk}
\end{equation}
there $\Rf$ is {\em reflection} matrix
defined as $\Rf \ket{n} = \ket{l-n-1}$. Such decomposition was used in a
{\em representation of quantum computation in phase space} with
{\em a discrete Wigner function} \cite{MPS,Paz}, and formally very similar
with ideas described below, but in applications with Lie algebras
matrix $\Rf$ in such products ``spoils'' some equations used in proof of
universality and so this representation is not used here.
\end{quote}

In a non-Hermitian case an universal set of elements may be based on
arbitrary complex matrices $M_k$ if the matrices generate full Lie algebra of
traceless matrices using sums and commutator. It is enough to
consider unitary gates
\begin{equation}
 G_k^\tau =e^{i(M_k+M_k^*)},\quad G_k^{\prime\tau} = e^{(M_k-M_k^*)\tau}.
\label{GG}
\end{equation}

Due to such a method it is possible to use the non-Hermitian matrix basis like
$U^k V^j$ without a special care. It is only necessary to prove, that all
such products may be generated using only commutators. It is not very
difficult, because the commutators are proportional to the products
$[U^a V^b,U^c V^d] = (\zeta^{-bc}-\zeta^{-ad})U^{a+c}V^{b+d}$ and it
is only necessary to be carefull with commuting elements \cite{VlaUNt}.

Here was only discussed an example with one system, but similar methods
may be used for the composition, using an analogue with Clifford algebras.
Let us denote
\begin{equation}
 \btau_1 = \op U, \quad
 \btau_2 = \zeta^{(l-1)/2} \op U \op V, \quad
 \btau_3 = \op V,
\end{equation}
where the complex multiplier $\zeta^{(l-1)/2}$ is used for normalization
$\btau_2^l = 1$. So here are analogues of all three Pauli matrices
with properties
\begin{equation}
 \btau_1 \btau_2 = \zeta \btau_2 \btau_1,\quad
 \btau_1 \btau_3 = \zeta \btau_3 \btau_1,\quad
 \btau_2 \btau_3 = \zeta \btau_3 \btau_2,\quad
 \btau_j^l = 1.
\label{taucom}
\end{equation}

Using these matrices it is possible to write an analogue of \Eq{defE}
\begin{eqnarray}
 \f_{2k} & = &
  {\underbrace{\btau_0\otimes\cdots\otimes \btau_0}_{n-k-1}\,}\otimes
 \btau_1\otimes\underbrace{\btau_3\otimes\cdots\otimes\btau_3}_k \, ,
 \nonumber\\
 \f_{2k+1} & = &
 {\underbrace{\btau_0\otimes\cdots\otimes \btau_0}_{n-k-1}\,}\otimes
 \btau_2\otimes\underbrace{\btau_3\otimes\cdots\otimes\btau_3}_k \, ,
 \label{defF}
\end{eqnarray}
where $k = 0,\ldots,n-1$ and $\btau_0$ is $l\times l$ unit matrix.

It can be checked directly, that the elements are satisfying to \Eq{zetcom}
and it is also possible to construct the full Lie algebra sl$(l^n,\C)$
using these $2n$ elements and so the construction described by
\Eq{GG} produces the set of universal quantum gates \cite{VlaUNt}.

It should be mentioned, that here again may be used the construction
with only one- and two-gates, if to consider the set with $2n$
elements
\begin{equation}
 \f_0,\quad \f_k\f_{k+1}^*\quad (k=0, \cdots, 2n-2)
\end{equation}
and to use two exponential formulas \Eq{GG}.

\medskip

It is also possible to consider the elements $\f_j$ \Eq{defF} from
point of view of Weyl quantization. Despite of the specific
form such elements may be directly derived from a general theory
if instead of the canonical {\em commutator form} \cite{WeylGQM},
i.e. $2n \times 2n$ symplectic matrix
\begin{equation}
  \Mat{rrrrr}{0&1&0&0&\ldots\\-1&0&0&0&\ddots\\
                   0&0&0&1&\ddots\\0&0&-1&0&\ddots\\
                 \vdots&\ddots&\ddots&\ddots&\ddots}
\label{CanCom}
\end{equation}
to use the special (non-canonical) form \cite{WeylClif}
\begin{equation}
 \Mat{rrrrrr}{0&1&1&1&1&\ldots\\-1&0&1&1&1&\ddots\\
 -1&-1&0&1&1&\ddots\\-1&-1&-1&0&1&\ddots\\-1&-1&-1&-1&0&\ddots\\
  \vdots&\ddots&\ddots&\ddots&\ddots&\ddots} .
\label{ClifCom}
\end{equation}
Here the elements of the commutator form $c_{ij} \in \{+1,-1,0\}$ correspond
to the relations $\f_i\f_j = \zeta^{c_{ij}}\f_j\f_i$.

\medskip

It should be mentioned what such ``$\zeta$-commuting'' elements are
also are quite common in the theory of {\em quantum algebras} \cite{Verb,Demid},
but this theory has a bit different prerequisites based on applications
of so called {\em R-matrix} for Yang-Baxter equation, the soliton
theory etc. \cite{Tur,FaddSol} and so the new constructions related
with the theory of quantum computations seem quite promising.

\section{Quantum computation in action}
\label{Sec:QC}

It was considered an approach to the quantum computation more similar with
an initial idea of universal physical operations \cite{FeySim,DeuTur,UnSim}.
Maybe such an approach more close to the ideas of {\em a quantum control}
\cite{SGRR02,ZL03}, because the universality was discussed without necessity of
the relation with ``traditional'' computing tasks. Such a distinction
may be a bit formal, but it should be emphasized what the quick growth of
an interest to the area of the quantum computation after about fifteen years
of a latent development was related with such a typically ``arithmetic'' task
as the factorization of numbers by a quantum computer using P. Shor algorithm
\cite{Shor} or quantum error correction codes \cite{KnillQEC,GottQEC,ShorQEC}.

In previous sections it was considered, how to exploit the whole set of
states and transformations using only some basic operations. For composition
of few quantum systems the spaces grow {\em exponentially} with respect to the
number of elements. It was shown that it is possible anyway to use only
Hamiltonians for transformations of each elements (one-gates) together with
Hamiltonians of {\em pairwise} interactions (two-gates) to construct an
universal set.  It is clear, that the number of such gates grows {\em
linearly} with the number of elements.

So, despite of an universality of such a set of gates, the number of gates in
a product used for a presentation of the {\em general} element of SU$(l^n)$
may be exponential. Due to it, not only the universal gates are necessary, but
also some special sets and constructions for a particular task.
Here is not discussed the general theory of quantum computations, and
currently there are lot of papers and monographs like \cite{BEZ,BDT}.
Only some selected point are mentioned below.

It is interesting to compare general ideas of construction of some new
mathematical structures for resolution of particular tasks. It was
already mentioned above, what abstract models, like {\em the algebraic
integer numbers} were constructed for research of some equation with
natural numbers. Such construction used some extension of
the idea of integer number \cite{FermPost}.

The theory of quantum computation uses similar ideas, because for the research
of computational problems, defined usually as some operations with numbers or
other finite models like Boolean algebras, are used some ``continuous''
algebraic extensions of such models. Even if technologically the large-scale
quantum computers, satisfying to all subtleties used in such abstract theories,
would never be build, they are already provide a great impact to the both
fields of pure mathematics and the quantum theory.

Similarly with the ideas of Lam\'e and Kummer about generalisation of natural
numbers, the quantum computer science instead of natural numbers $0,1,\cdots$
or Boolean values $0,1$ (\{{\bf false},{\bf true}\}) uses formal series like
$\alpha_0\ket{0}+\alpha_1\ket{1}$, $\alpha_k \in \C$.

Instead of a discrete finite set of functions with integer or boolean
values it has used a continuous group, that also may be represented as
a formal series. To show it, let us express usual ``classical'' functions
with notations more habitual in the quantum information science
(cf \cite{Merm02,Merm03}).

It was already mentioned, that $\bsig_x$ corresponds to a {\sf NOT} gate
in the usual computation. Let us denote a matrix with all zeros except one
unit in the position $a_{ij}$ as $\ktbr{i}{j}$, so the {\sf NOT} gate may be
represented as the sum of two such matrices $\ktbr 01 + \ktbr 10$.
Similarly an arbitrary function on $\Z_l = \{0,\ldots,l-1\}$ may be written as
$
M_f = \sum_{k=0}^{l-1}\ktbr{f(k)}{k}.
$
\begin{quote}
{\em Note:} Only for the reversible function $f$ the matrix $M_f$ is orthogonal,
but here is a simple trick to associate a reversible function $F$ with
any irreversible one, the function is defined on pairs of numbers
$F\colon (x,y) \mapsto (x,f(x)+y \mod l)$ and so produces for a pair
$(x,0)$ the pair of values $(x,f(x))$ (this idea is widely used in the
theory of quantum computation, there it is denoted as
$\ket{x}\ket{y} \mapsto \ket{x}\ket{f(x)+y \mod l}$ \cite{BEZ,VBE,revis}).
Formally such a function may be considered as defined on $\Z_{l^2}$.
\end{quote}

A complex matrix may be expressed as the similar sum
$
U = \sum_{j,k=0}^{l-1}U_{jk}\ktbr{j}{k}
$
and unitary matrices are analogue of reversible functions, because
for a reversible function $M_f$ corresponds to an unitary (orthogonal) matrix
with $l$ units and $l^2-l$ zeros.

In such a correspondence with classical computations Weyl pair of matrices $U,V$
also have interesting properties. Say elements of Hilbert space representing
natural numbers, i.e. {\em the computational basis} $\ket{0},\ket{1},\cdots,
\ket{l-1}$, are eigenvectors of the matrix $V$ with eigenvalues $\zeta^k$:
$V\ket{k}=\zeta^k \ket{k}$ and the matrix $U$ corresponds to a cyclic shift of
the elements $U\ket{k} = \ket{k+1 \mod l}$.

Eigenvectors of the matrix $U$ may be expressed as
\begin{equation}
\widetilde{\ket{k}} = \sum_{j=0}^{l-1} \zeta^{kj}\ket{j},
\quad k=0,\ldots,l-1,\quad \zeta^{kj} = \exp(2\pi i\,kj/l),
\end{equation}
and have an analogue with a basis in momentum space.
Transition between the computational basis $\ket{k}$ and the momentum basis
$\widetilde{\ket{k}}$ may be represented by a matrix $F$ with indexes
$F_{kj}=\zeta^{kj}$, i.e.
\begin{equation}
 F = \Mat{ccccc}{1&1&1&\ldots&1\\1&\zeta&\zeta^2&\ldots&\zeta^{l-1} \\
 1&\zeta^2&\zeta^4&\ldots&\zeta^{2l-2}\\
 \vdots&\vdots&\vdots&\ddots&\vdots\\
 1&\zeta^{l-1}&\zeta^{2l-2}&\ldots&\zeta^{(l-1)(l-1)}}\!.
\label{QFT}
\end{equation}
The operator $F$ is called {\em the discrete (or quantum) Fourier transform}
and the most principal quantum algorithms are just based on application
of such transform \cite{JozF} and possibility of fast
implementation using the special set of quantum gates \cite{JozE}.
It is some demonstration of the notion, that not only an universal
set is necessary, but the special gates for the fast implementations
of the specific transformations.

Another natural area for application of Weyl pair $U,V$ is the quantum
error correction codes, there the products and bases like $U^k V^j$ can
be directly used for construction of such codes \cite{KnillQEC,GottQEC}.
Such situation maybe not so unexpected, because the theory of quantum error
correction codes ``borrows'' some part from the classical one \cite{EO,GF4}
and so was related with an extensive branch of the discrete mathematics
\cite{Pack}. Idea of {\em stabilizer codes} \cite{GF4,GottPhD} has connected
that area with specific commuting relation and, finally, with Weyl pair
\cite{GottQEC,GottPhD,AK}.

So specific constructions like Galois fields \cite{GF4,AK} and algebraic
numbers may be naturally interlaced with the quantum mechanical ideas.  It is
relevant not only to the theory of quantum error correction codes, for
example similar ideas may be applied to construction of mutually unbiased
bases used for the security of quantum communications \cite{Boykin} and the
theory of quantum measurements \cite{mub,SymSet}. For such applications there
are also interesting analogues with the discrete Wigner function defined in
\cite{MPS,Paz} and briefly mentioned in \Sec{Weyl} in relation with
alternative decomposition \Eq{Ajk}.

\end{document}